\documentclass[aps,pre,showpacs,showkeys,twocolumn,groupedaddress,amsmath]{revtex4-1}

\usepackage{graphicx}
\usepackage{epsfig}
\usepackage{epstopdf}
\usepackage{bm}
\usepackage{nicefrac}
\usepackage{color}

\begin{document}


\title{Motility of active fluid drops on surfaces}


\author{Diana Khoromskaia}
\email[]{D.Khoromskaia@warwick.ac.uk}
\author{Gareth P. Alexander}
\affiliation{Department of Physics and Centre for Complexity Science, University of Warwick, Coventry CV4 7AL, United Kingdom}


\date{\today}

\begin{abstract}
Drops of active liquid crystal have recently shown the ability to self-propel, which was associated with topological defects in the orientation of active filaments [Sanchez {\em et al.}, Nature {\bf 491}, 431 (2013)]. 
Here, we study the onset and different aspects of motility of a three-dimensional drop of active fluid on a planar surface. We analyse theoretically how motility is affected by orientation profiles with defects of various types and locations, by the shape of the drop, and by surface friction at the substrate. In the scope of a thin drop approximation, we derive exact expressions for the flow in the drop that is generated by a given orientation profile. The flow has a natural decomposition into terms that depend entirely on the geometrical properties of the orientation profile, i.e.\ its bend and splay, and a term coupling the orientation to the shape of the drop. We find that asymmetric splay or bend generates a directed bulk flow and enables the drop to move, with maximal speeds achieved when the splay or bend is induced by a topological defect in the interior of the drop. In motile drops the direction and speed of self-propulsion is controlled by friction at the substrate.

\end{abstract}


\maketitle

\section{Introduction}
\label{Sec:Introduction}

Active fluids are non-equilibrium materials composed of elongated units that constantly consume energy to move themselves forward or to stir their surrounding medium \cite{Sanchez:2013gt,Marchetti:2013bp,Ramaswamy:2010bf}. Living systems that can be described as active fluids are often found in spatial confinement, like suspensions of swimming bacteria in porous media \cite{Rusconi:2014gd} or the cytoskeleton \cite{Julicher:2007fy} of a cell enclosed by its membrane. Experiments with active fluids confined to droplets revealed a myriad of novel features, such as spontaneous symmetry breaking \cite{AbuShah:2014jx}, self-organised defect structures \cite{1997Natur.389..305N,Sanchez:2013gt,Keber:2014fh}, coherent large-scale flows \cite{Wioland:2013jm,Lushi:2014fn,2012PNAS..10911705P,Kumar:2014kv} and even self-sustained motility \cite{Sanchez:2013gt}. A full theoretical understanding of the interplay of activity with total geometrical confinement is still lacking, yet it could not only help to unveil the basic fluid dynamical  processes underlying cell motility and cell division, but also allow to tune the macroscopic behaviour of biomimetic droplets \cite{Sanchez:2013gt,Keber:2014fh}, for instance by controlling the orientation of active filaments. 

Previous theoretical work on active droplets has focused on two-dimensional geometries or spherical droplets immersed in an other fluid. The active fluids that these models typically refer to are aqueous suspensions of cytoskeletal filaments together with their associated motors, in particular actin filaments with myosin motors as a material having polar symmetry or microtubules with kinesin motor clusters \cite{Sanchez:2013gt} as one having nematic symmetry, respectively.  In a circular domain an active fluid can self-organise into a stable circulating state at high enough activity \cite{Woodhouse:2012cl}, which is related to cytoplasmic streaming in plant cells. A similar rotating spiral, albeit with a counterrotating boundary layer and varying cell orientation, was observed in a quasi-two-dimensional circular suspension of swimming bacteria \cite{Wioland:2013jm}. Hydrodynamic interactions among the bacteria and with the confining wall were identified to be crucial to drive this self-organised pattern \cite{Lushi:2014fn}. Nematic two-dimensional droplets showed elongation, spontaneous division and motility as a result of the interplay of defects, geometry and activity \cite{Giomi:2014ce}.  Finally, in a numerical study of polar active droplets of initially circular (2D) or spherical (3D) shape, surrounded by a passive fluid, increasing activity led to spontaneous symmetry breaking accompanied by self-propulsion and deformation \cite{2012PNAS..10912381T,Marth:2015eb}. These findings were underpinned theoretically by the analysis of a droplet of fixed shape and with an imposed splayed orientation field \cite{AWhitfield:2014in}. A related class of models is concerned with the cellular actin cortex as an active gel and provided insight into activity-driven deformation and migration of cells \cite{Turlier:2014hq,Mayer:2010kt,Hawkins:2011kl}. 

The behaviour of a three-dimensional active drop placed on a flat surface \cite{Joanny:2012dx,2015NatCo...6E5420T,2010BpJ....98.1408H} has been hitherto scarcely investigated. However, it is important to understand this setup in order to clarify the role of active flows in motile cells on  substrates \cite{Keren:2009gn} and, in a broader fluid dynamical perpective, to characterise novel phenomena not found in the well-studied passive fluid droplets on surfaces \cite{1997RvMP...69..931O}. Topological defects in the orientational order, which are points of undefined orientation \cite{deGennes_Prost}, emerge spontaneously in active systems \cite{Sanchez:2013gt}. Due to strong gradients of active stresses in their vicinity, defects can act as sources for large-scale flows which, in combination with the confined shape, can drive macroscopic motion of a drop of active fluid along a surface. To analyse which types of orientation profiles with defects lead to three-dimensional flows that drive a drop from within, and how surface friction at the substrate and the shape of the drop affect these flows, is the purpose of this work. 

A first theoretical analysis of active drops on surfaces focused on spreading laws and stationary shapes for several polarisation fields with high symmetry \cite{Joanny:2012dx}, accounting for a no-slip boundary only. In a recent numerical study of a drop of polar active fluid various biologically relevant shapes and motile steady states were obtained \cite{2015NatCo...6E5420T}. The computation included both actin treadmilling and a variable effective friction representing the focal adhesions as mechanisms that are crucial for cell motility on a substrate. In another computational study realistic three-dimensional shapes of motile cells were generated using a hydrodynamic model \cite{2010BpJ....98.1408H}, although without including the polarisation field of active filaments.

In this paper we consider generic flow profiles that arise in three-dimensional droplets on planar substrates with a variety of polarisation fields associated to different types of defects. We derive analytical expressions for the flow field resulting from a specified polarisation and varying friction at the substrate. This leads to a geometric description of the flow, with a natural separation of splay and bend into horizontal and vertical components, of which only the latter depends on the shape of the drop. We find that self-propulsion of the drop is enabled by asymmetric orientation fields and is fastest when the splay or bend is induced by a defect approximately mid-way between the centre and boundary of the drop, for instance an aster or a vortex. When the substrate friction is negligible the interior flow in the drop is rotational and there is no propulsion. Maximal propulsion is achieved when the friction is greatest. 
The shape of the drop is free to deform as a result of the active flows. We qualitatively describe the deviation from an initial spherical cap shape for two examples. For the case of an aster defect in the centre of an axisymmetric drop we derive the stationary shape of an active drop. 

The paper is structured as follows. In Sec.\ \ref{Sec:Model} the model for a thin drop of active fluid is derived, including the equations for the flow (\ref{Subsec:Equations}), the most general form of a polarisation field subject to tangential anchoring (\ref{Subsec:ActiveStress}), and the approximation of equations (\ref{Subsec:SeparationScales}) and boundary conditions (\ref{Subsec:BoundaryConditions}) under the assumption of separable length scales. The exact general solution for the flow field is presented at the start of Sec.\ \ref{Sec:Results}. In Sec.\ \ref{Subsec:PolarisationDependence} the structure of the horizontal flow is explained and illustrated by examples of various orientation fields. In Sec. \ref{Subsec:FrictionDependence} the implications of the surface friction on motility are discussed and Sec.\ \ref{Subsec:ShapeDeformation} is concerned with shape deformations and stationary shapes. Finally, a discussion of the results and their relation to experiments is provided in Section \ref{Sec:Discussion}.

\begin{figure}[]
\includegraphics[width=0.48\textwidth]{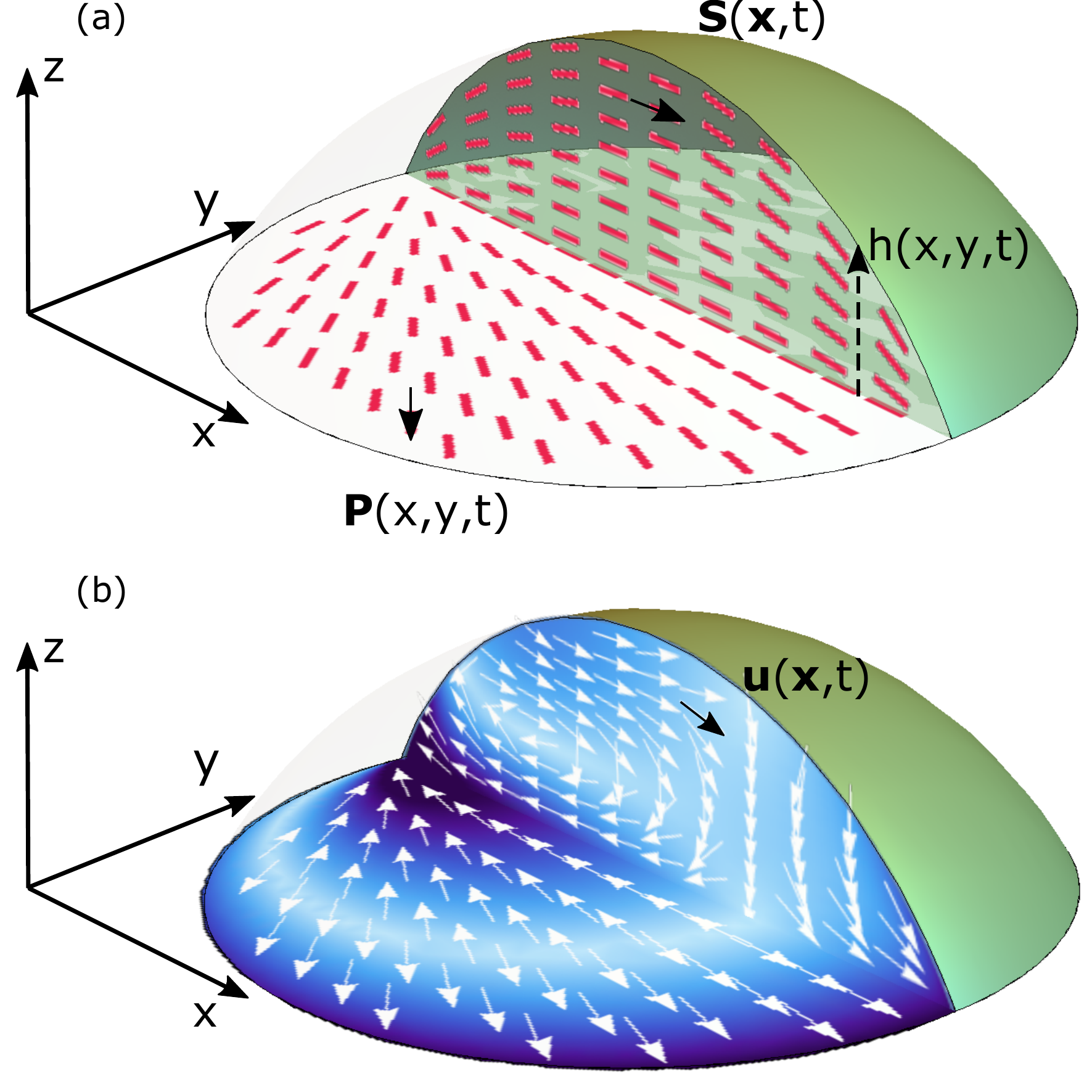}
\caption{\label{fig:Setup} (a): Schematic view of a droplet of active fluid with filaments oriented according to the field $\bm{S}$, which has the two-dimensional projection $\bm{P}$. The filaments are anchored parallel to the bounding surface, which is given by its height function $h$. (b): The active stresses induced by the filaments generate a flow $\bm{u}$. }
\end{figure}

\section{Model}
\label{Sec:Model}

 We consider a three-dimensional drop of active fluid, illustrated in Fig.\ \ref{fig:Setup}, with mixed boundary conditions: a flat, rigid substrate providing frictional dissipation underneath and air or a fluid of much lower viscosity with zero tangential stress above the drop, allowing the shape to deform. The active fluid in our model is either an active liquid crystal \cite{AditiSimha:2002eg} or an active polar gel \cite{Kruse:2005fy}, representing a suspension of active filaments in nematic or polar long-range orientational order, respectively. We study how the onset and different features of motility of such drops are controlled by spatial variations and defects in the orientation field, the shape of the drop, and the surface friction at the subtrate. To address these questions we derive a simple hydrodynamic model, which is built on two basic assumptions. 

Active fluids are described theoretically in terms of generalised hydrodynamic equations that couple the dynamics of the flow and the orientation field and can be derived by coarse-graining a suspension of active particles \cite{AditiSimha:2002eg,Simha:2002gd,Hatwalne:2004dy} or by modelling the cytoskeleton in terms of a viscoelastic gel driven out of equilibrium \cite{Kruse:2004il,Kruse:2005fy,2015NatPh..11..111P}. 
The equations are a coupled non-linear system for both the orientation field of the active filaments and the velocity field of the active fluid. A simplification can be obtained for dynamic steady states by assuming that the alignment is static and close to the energy-minimising configurations of passive liquid crystals. That given, one need only solve for the active fluid flow. 
We impose tangential anchoring at both the base of the drop and the free surface, while at the contact line the orientation is free to point in any direction. This choice of anchoring is motivated among others by experimental setups where the filaments adhere to the boundary layer in large enough droplets \cite{Sanchez:2013gt,Keber:2014fh}. Also, in the lamellipodia of crawling cells actin filaments lie tangential to the substrate and push perpendicularly against the leading edge \cite{Yam:2007kk}.  
 
 The second assumption is that the average height of the drop is much smaller than its width and we can separate the two length scales to simplify the equations for the flow field, as is typically done for thin fluid films \cite{Sankararaman:2009bx,Joanny:2012dx}. For this purpose we choose the drop to be initially in the shape of a flat spherical cap with radius $r_d$ and height function given by $z=h(x,y)=\sqrt{1-x^2-y^2}-0.5$ (see Fig.\ \ref{fig:Setup}(a)). However, the shape is free to deform due to the active flows generated in the drop and these deformations will also change the orientation field indirectly, through its anchoring to the surface.  
 
 These assumptions allow us to find an exact solution for the flow field in a drop of a given shape and with a prescribed orientation profile. Its structure turns out to be guided by both the familiar instabilities towards bend and splay in active systems \cite{AditiSimha:2002eg} and the coupling of the orientation to gradients in the height profile, which has previously been studied in the context of spreading of drops \cite{Joanny:2012dx}. The exact formula for the flow allows us to address a variety of aspects of drop motility in an analytical way. Our model reveals that asymmetric spatial variations in the orientation are key to enable drop propulsion and explains why friction at the substrate is essential. We also find that topological defects can both drive drop motility and also induce singular vertical flows that could lead to the growth of protrusions at the site of the defect.

\subsection{Governing equations}
\label{Subsec:Equations}

Here, we only consider the dynamics of the flow field $\bm{u}(\bm{x},t)=(u,v,w)$ and the pressure $p(\bm{x},t)$ of the suspension, which comprises both the active particles and the solvent fluid as a whole, and the resulting variations in the height profile of the drop, given by $z=h(x,y,t)$. The hydrodynamic effect of the active filaments on the suspension is an active stress, whose gradient drives the flow. To first order in a gradient expansion the active stress tensor for nematic and polar suspensions is indistinguishable \cite{AditiSimha:2002eg,Joanny:2012dx}, which is applicable here since we are interested in flows that are of large scale compared to the size of a filament. As the systems described here are in the low Reynolds number regime, the dynamics of the suspension is given by the generalised Stokes and continuity equations \cite{Joanny:2012dx}
\begin{eqnarray}
\label{eq:NavierStokesEquation}
- \nabla p + \mu \Delta \bm{u} + \nabla \cdot (\bm{\sigma}^a+ \bm{\sigma}^e) & = & 0 ,\\
\quad \nabla \cdot \bm{u} & =  & 0 \,, \label{eq:ContinuityEquation}
\end{eqnarray}
where $\rho$ is the density, $\mu$ the viscosity, $\bm{\sigma}^e$ the Ericksen stress of passive liquid crystals, and $\bm{\sigma}^a$ the active stress. The boundary conditions for equations \eqref{eq:NavierStokesEquation} and \eqref{eq:ContinuityEquation} are specified below, in Sec.\ \ref{Subsec:BoundaryConditions}.

\subsection{Polarisation field and active stress}
\label{Subsec:ActiveStress}

The polarisation is taken to be tangential at both bounding surfaces, the flat base in contact with the substrate and the free upper surface of the droplet, $z=h(x,y,t)$. If the polarisation on the base ($z=0$) is $(P_x, P_y,0)$, with $|\bm{P}|=1$, then this can be matched onto a tangential polarisation on the upper surface by setting
\begin{equation}\label{eq:Zcomponent}
P_z = (P_x \partial_x h + P_y \partial_y h )/ \sqrt{1+ (\partial_x h)^2 + (\partial_y h)^2} \,\,.
\end{equation}
The resulting vector field lies tangential to the surface with the unit outward normal $\bm{n}_h =(-\partial_x h, -\partial_y h, 1)^{\top}/ \sqrt{1+(\partial_x h)^2+(\partial_y h)^2}$. 
Finally, the three-dimensional polarisation field $\bm{S}(\bm{x})$ is constructed by smoothly interpolating between the two surfaces using the factor $z/h$ in the $z$-component, 
\begin{equation}\label{eq:Polarisation3D}
\bm{S}(\bm{x}) = \frac{1}{n(\bm{x})}\left(
\begin{array}{c}
h P_x\\
h P_y\\
z P_z\\
\end{array}
\right)\,, 
\end{equation}
and normalising with $n(\bm{x})= \sqrt{h^2+ z^2 P_z ^2}$. This is the most general form of a polarisation field that satisfies the tangential anchoring condition and is consistent with the assumption of a thin drop, with small variations of the polarisation in the $z$-direction.

The active stress tensor is defined as \cite{AditiSimha:2002eg}
\begin{equation}\label{eq:ActiveStress}
\sigma^a _{ij}= -\sigma_0 \left(S_i S_j - \frac{\delta_{ij}}{3}\right) \,\, ,
\end{equation}
where the constant $\sigma_0$ sets the strength of the activity and whether the stress is contractile ($\sigma_0 <0$) or extensile ($\sigma_0 >0$), that is whether the active particles drive fluid in or out along their axis of orientation, respectively.  

\subsection{Separation of length scales} 
\label{Subsec:SeparationScales}

We can exploit the geometry of the drop to simplify equation (\ref{eq:NavierStokesEquation}). We assume that its horizontal extension, characterised by the length scale $L$, is much larger than its average height $h_0$. Therefore, as we are interested in long-wavelength phenomena, the variations of the flow in $x$ and $y$ directions are much more gradual than in $z$ direction. We define the small parameter $\varepsilon = \frac{h_0}{L} \ll  1$ and perform an expansion of the governing equations, similar to the approach for thin films \cite{1997RvMP...69..931O,Sankararaman:2009bx,Joanny:2012dx}.

We make the coordinates dimensionless by scaling according to $x \to \frac{\varepsilon x}{h_0}$, $y \to  \frac{\varepsilon y}{h_0}$, $z \to \frac{z}{h_0}$, and likewise $h \to \frac{h}{h_0}$. Compared to the height the extension of the drop is similar in both horizontal directions, therefore we treat the $x$ and the $y$ directions analogously. With a characteristic velocity $U_0$ we scale $u \to  \frac{u}{U_0}$ and $v \to \frac{v}{U_0}$. For the vertical flow component the incompressibility condition \eqref{eq:ContinuityEquation} requires $w \to \frac{w}{\varepsilon U_0}$. Finally, the time scale of the flow is set by $L/U_0$, and therefore $t \to \frac{\varepsilon U_0 t}{h_0}$.

When we expand equation \eqref{eq:NavierStokesEquation} in $\varepsilon$ then, in the $x$ component for instance, the leading order term is $\sim \partial_z^2 u $, as expected for a thin film  \cite{1997RvMP...69..931O}, while the contributions due to pressure and active stress are of sub-leading order. Therefore, to retain the effect of activity these terms are scaled as
\begin{equation}\label{eq:PSigmaScaling}
p \to \frac{\varepsilon h_0}{\mu U_0} p \quad \textrm{and} \quad
\sigma_0 \to \frac{\varepsilon h_0}{\mu U_0} \sigma_0  .
\end{equation}
The field  $\bm{P}$ is normalised, so its components are $P_x, P_y \sim  O(1)$. It follows that the $z$-component \eqref{eq:Zcomponent} should be rescaled as $P_z \to  P_z / \varepsilon$ and the normalisation of $\bm{S}$ becomes $n(\bm{x}) \approx h_0 h(x,y)$. With these scalings the dominant terms of the gradient of the active stress can be written as $(\nabla \cdot \sigma^a)_{\perp} = \frac{\mu U_0}{h_0^2} \bm{f}_{\perp}^a$, 
 in terms of an effective active force
 \begin{equation}\label{eq:ActiveForce}
\bm{f}_{\perp}^a = -\sigma_0 \left(\!  \bm{P} \! \left(\nabla_{\!\perp}\! \cdot \bm{P} +\frac{1}{h}\bm{P}\cdot \nabla_{\!\perp} h \!\right) \!+ \left(\bm{P} \! \cdot \! \nabla_{\!\perp} \right) \! \bm{P} \right) ,
 \end{equation}
 which only depends on the horizontal position $\bm{x}_{\perp}=(x,y)$. We find that $ \bm{f}_{\perp}^a$ has a natural decomposition into terms associated with the geometric properties of $\bm{P}$, that is its splay and bend, and a term that couples $\bm{P}$ to gradients in the height profile of the drop. This structure is inherited by the resulting flow field, which is driven by  $ \bm{f}_{\perp}^a$, and will be discussed in detail in Sec.\ \ref{Subsec:PolarisationDependence}.

The tensor $\bm{\sigma}^e$ in \eqref{eq:NavierStokesEquation} accounts for standard nematic elasticity \cite{deGennes_Prost}. Assuming one elastic constant $K$, the leading order term in the elastic distortion free energy density is the one associated with splay, $\frac{K}{2}\left(\nabla \cdot \bm{S}\right)^2 \sim O(\varepsilon^2)$. Thus, in relation to the active stresses we can omit $\bm{\sigma}^e$. In addition, the orientation profiles we consider differ from the ones that minimise elastic energy for passive liquid crystals only by a small vertical tilt, suggesting that flows due to elasticity should be small compared to active flows. 

To leading order in $\varepsilon$ equation \eqref{eq:NavierStokesEquation} becomes
\begin{eqnarray}\label{eq:ThinFilmEquation1}
\partial_z ^2 \bm{u}_{\perp} &= & \nabla_{\perp} p  - \bm{f}^a_{\perp} ,\\
0 & =& \partial_z p \,, \label{eq:ThinFilmEquation2}
\end{eqnarray}
where $\bm{u}_{\perp}= (u,v)$ and $\nabla_{\perp} = (\partial_x, \partial_y)$, while the continuity equation remains unchanged.

\subsection{Boundary conditions}
\label{Subsec:BoundaryConditions}

In this section we first state the boundary conditions in terms of unscaled variables, before transforming to dimensionless variables and expanding in the small parameter $\varepsilon$. At the base we allow for partial slip with linear friction, such that the shear stress acting on the fluid is proportional to the fluid velocity at the boundary,
\begin{equation}\label{eq:LinearFrictionBC}
\bm{n}^\top_0 \bm{T} \hat{\bm{e}}_i = - \frac{1}{\xi}u_i \quad\textrm{for } i=x,y \,\,.
\end{equation}
Here $\xi$ is the inverse friction coefficient, so that $\xi=0$ corresponds to the no-slip condition. The full stress tensor of the fluid is given by
\begin{equation}\label{eq:FullStressTensor}
T_{ij} = -p \delta_{ij} + \mu \left( \partial_j u_i + \partial_i u_j \right) + \sigma^a_{ij} + \sigma^e_{ij},
\end{equation}
and $\bm{n}_0 = -  \hat{\bm{e}}_z $ is the outward normal to the lower bounding surface. After making the inverse friction coefficient dimensionless by scaling it as $\xi \to \frac{\mu \xi}{h_0} $, for small $\varepsilon$ condition \eqref{eq:LinearFrictionBC} becomes
 \begin{equation}\label{eq:BCSlip}
\bm{u}_{\perp}(z=0) = - \xi \partial_z \bm{u}_{\perp}\bigg| _{z=0}\,,
 \end{equation}
now written in dimensionless quantities. The substrate is non-permeable and therefore the $z$-component of the flow must vanish there, $w|_{z=0} =0$ .

At the free boundary the stress component tangent to the surface should vanish, 
\begin{equation}
 \bm{n}_h^\top \bm{T} -( \bm{n}_h^\top \bm{T}\bm{n}_h)  \bm{n}_h =\bm{0},
\end{equation}
which for small $\varepsilon$ and in dimensionless variables simplifies to
\begin{equation}\label{eq:BCUpper}
 \partial_z \bm{u}_{\perp}\big| _{z=h} =\bm{0}.
\end{equation}

The normal stress at the upper boundary is given by the mean curvature of the surface $\nabla \cdot \bm{n}_h$, the surface tension $\gamma$ and the pressure $p_0$ in the surrounding medium,
\begin{equation}\label{eq:YoungLaplace}
p =  p_0 +\frac{\sigma_0}{3} -\varepsilon^3 \frac{h_0 \gamma}{\mu U_0} (\partial_x^2 h + \partial_y ^2 h) + \mathcal{O}(\varepsilon^2)\,\,,
\end{equation}
which in the absense of activity reduces to the Young-Laplace equation. In \eqref{eq:YoungLaplace} only $\gamma$ still carries dimensions. The surface tension term is three orders of magnitude smaller than the leading term. If we wanted to retain surface tension effects, the coefficient had to scale like $\gamma \sim \varepsilon^{-3}$. However, earlier we identified that the active stress coefficient scales as $\sigma_0 \sim \varepsilon^{-1}$ (cf.\ eq.\ \eqref{eq:PSigmaScaling}). Because here we are interested in active effects primarily, we will neglect the surface tension term. Then the boundary condition \eqref{eq:YoungLaplace} becomes $p|_{z=h} = p_0 + \frac{\sigma_0}{3}$. 

The upper bounding surface is free to deform and its time evolution is given by the flow component normal to the surface
\begin{equation}\label{eq:BCkinematic1}
\partial_t h = \bm{u}\cdot \bm{n}_h = w- u \partial_x h - v\partial_y h\,\,.
\end{equation}
This kinematic boundary condition can be expressed in terms of $u$ and $v$ only, if we integrate the continuity equation (with $w(0)=0$) and use Reynolds' theorem.
The height profile  then evolves according to
\begin{equation}\label{eq:BCkinematic2}
\partial_t h = - \nabla_{\perp} \cdot \left(\int\limits_{0}^{h} \bm{u}_{\perp} \textrm{d}z \right)  \,\,.
\end{equation}

\begin{figure*}[]
\includegraphics[width=\textwidth]{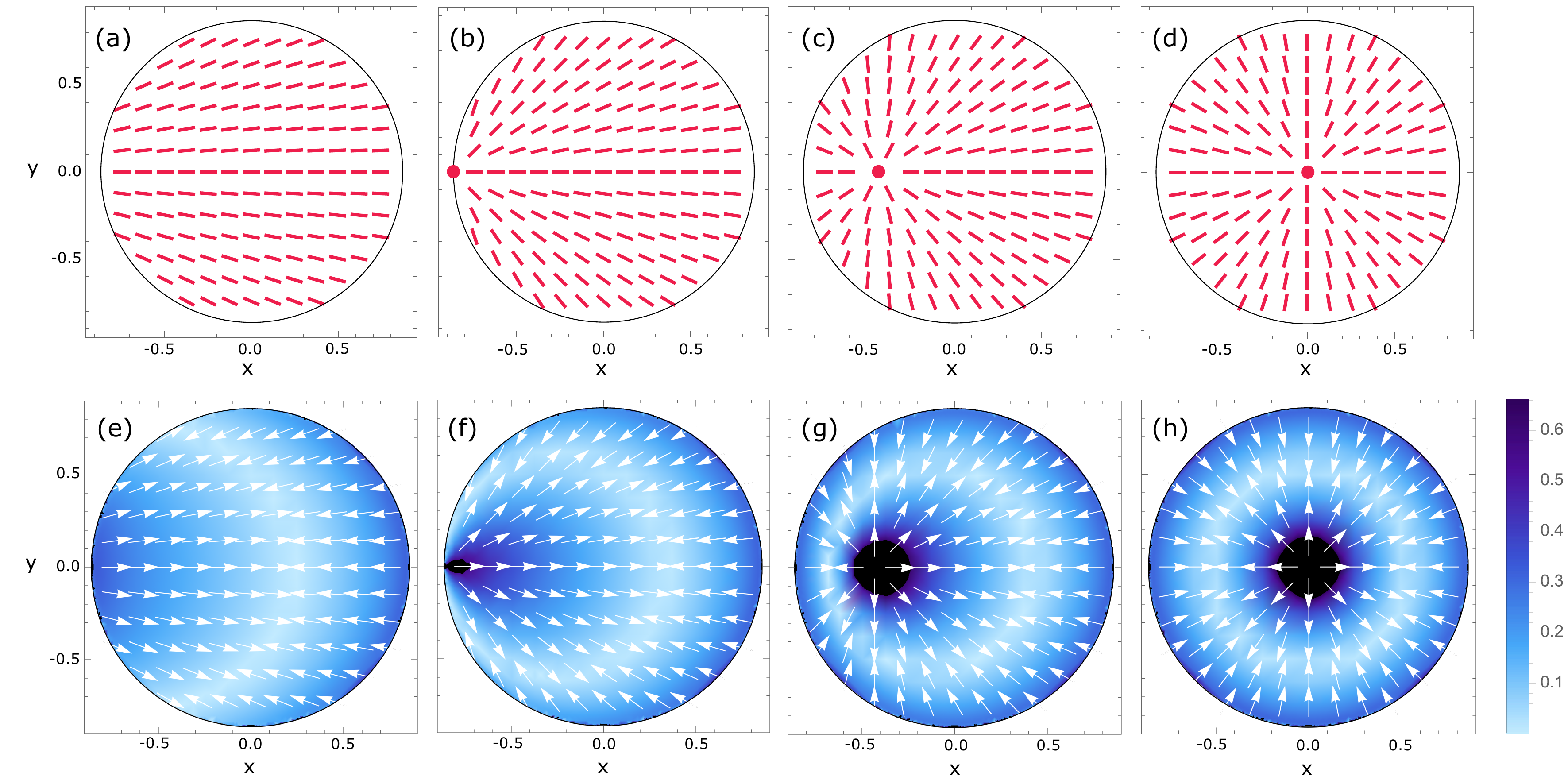}
\caption{\label{fig:AsterDefect} (a)-(d): Top view of drops with a splayed polarisation field ($a=0, b=1$) due to a defect at $r_0=-2r_d, -r_d, -\frac{1}{2}r_d, 0$. (e)-(h): Plot of the resulting flow field \eqref{eq:FlowDecomposition} at $z=0.01$, for a contractile drop with no slip ($\sigma_0=-1,\, \xi=0$). The flow is  aligned with the polarisation, but changes sign along a line where the splay of $\bm{P}$ balances the coupling term \eqref{eq:PHcoupling}. Here and in the following plots red lines represent the polarisation field, and the flow field is decomposed into direction (white arrows) and magnitude (colour coded, scaled). At a defect the flow typically diverges and is cut off in the display (black region).}
\end{figure*}
\begin{figure*}[]
\includegraphics[width=\textwidth]{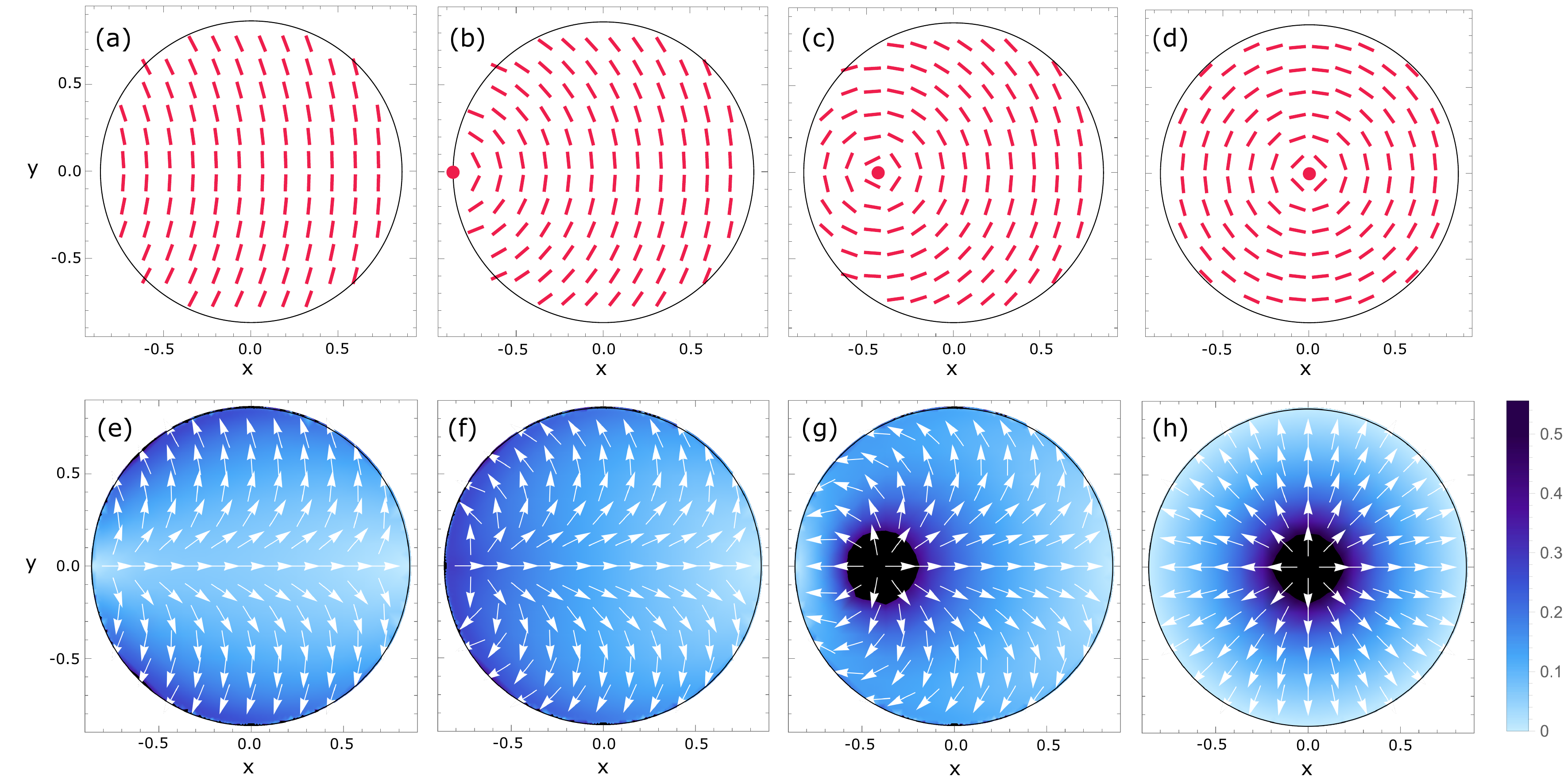}
\caption{\label{fig:VortexDefect} (a)-(d): Top view of drops with a bent polarisation field ($a=\frac{\pi}{2}, b=1$) due to a defect at $r_0=-2r_d, -r_d, -\frac{1}{2}r_d, 0$. (e)-(h): Plot of the resulting flow field \eqref{eq:FlowDecomposition} at $z=0.01$, for an extensile drop with no slip ($\sigma_0=1,\,\xi=0$). The flow is perpedicular to the polarisation in the bulk, but in (e)-(g) it alignes with the polarisation close to the boundary due to the coupling term \eqref{eq:PHcoupling}. In the axisymmetric case, (d) and (h), the coupling term vanishes and the flow points radially outwards.}
\end{figure*}

\section{Results}\label{Sec:Results}

We solve the equations \eqref{eq:ThinFilmEquation1} and \eqref{eq:ContinuityEquation} for the instantaneous flow field $\bm{u}$ resulting from a particular orientation field $\bm{P}$ and shape $h$. From \eqref{eq:ThinFilmEquation2} we find, that the pressure is constant to first approximation, $p=p_0+\frac{\sigma_0}{3}$, whereby the surface tension was neglected. Then, integrating equation \eqref{eq:ThinFilmEquation1} twice gives an expression for the horizontal flow components
\begin{equation}\label{eq:FlowDecomposition}
\begin{split}
\bm{u}_{\perp} & = \sigma_0 \!\left(\frac{z^2}{2}+ h \left(\xi -z\right)\right) \\ 
& \quad\, \times  \!\!\left(  \bm{P} \! \left(\nabla_{\!\perp}\! \cdot \bm{P} +\frac{1}{h}\bm{P}\cdot \nabla_{\!\perp}h \right) + \left(\bm{P} \! \cdot \! \nabla_{\!\perp} \right) \! \bm{P} \right)\\
&= -\left( \frac{z^2}{2}+ h \left(\xi -z\right)\right) \bm{f}_{\perp}^a\,.
\end{split}
\end{equation}
The $\bm{x}_{\perp}$-dependence of $\bm{u}_{\perp}$ is primarily determined by the effective active force $\bm{f}_{\perp}^a$ ( see eq.\ \eqref{eq:ActiveForce}), and can be decomposed into three parts, each with a clear interpretation in terms of spatial variations in $\bm{P}$ and $h$ (see Sec.\ \ref{Subsec:PolarisationDependence}). 
 The vertical flow component follows from the continuity equation \eqref{eq:ContinuityEquation},
\begin{equation}\label{eq:FlowVertical}
w = \frac{z^3}{6} \nabla_{\!\perp}\! \cdot \bm{f}_{\perp}^a -\left(\frac{z^2}{2} -\xi z \right) \nabla_{\!\perp}\! \cdot \left(h \bm{f}_{\perp}^a \right)\,,
\end{equation}
from which it is obvious that the three-dimensional nature of the flow will be most prominent in regions of strongly varying $ \bm{f}_{\perp}^a$, for instance at topological defects in $\bm{P}$. Fig.\ \ref{fig:Setup}(b) provides an insight into one example of the full three-dimensional flow field $(\bm{u}_{\perp},w)$, however in the following we restrict the visual presentation to two-dimensional cross-sections. 

We illustrate the properties of the flow in the drop on examples of polarisation fields that are generated by defects of different topological strength and are given in the form
\begin{equation}\label{eq:PolarisationField}
 \bm{P}=\Big(\cos \left(a+b\theta(r_0)\right), \sin \left(a+b\theta(r_0)\right)\Big)^{\top},
\end{equation}
where $b$ is the strength of the defect and $a$ controls the shape of the field around it, for example whether it is an aster or a vortex for strength +1 (see Fig.\ \ref{fig:AsterDefect}(d) and Fig.\ \ref{fig:VortexDefect}(d)). The angle $\theta(r_0)=\arctan \left( \frac{y}{x-r_0}\right)$ depends on the position $r_0$ of the defect along the $x$-axis, such that the defect can be located inside or outside of the drop (see Fig.\ \ref{fig:AsterDefect}(a) and (c)). We remind the reader that all results are presented in terms of dimensionless quantities introduced in Sec.\ \ref{Subsec:SeparationScales}.

\subsection{Splay and bend drive directed large-scale flows in the drop}\label{Subsec:PolarisationDependence}

Writing the effective active force, and thus the flow field, in a coordinate free form (see eq.\ \eqref{eq:FlowDecomposition}) allows us to interpret the contributions to the flow in terms of the geometrical properties of $\bm{P}$ and $h$. The first $\bm{P}$-dependent term in \eqref{eq:FlowDecomposition}  produces flow in direction of the polarisation field and is proportional to the sum of its splay, $\nabla_{\!\perp}\! \cdot \bm{P}$, and the term
\begin{equation}\label{eq:PHcoupling}
U_h = \frac{1}{h}\bm{P}\cdot \nabla_{\!\perp} h\,\,,
\end{equation}
which couples the polarisation of the filaments to the shape of the drop and accounts for the splay into the third dimension. Since term \eqref{eq:PHcoupling} is large in the vicinity of the contact line, where $h \to 0$, it dominates the flow at the boundary and aligns it there with the polarisation field. The second contribution, $(\bm{P} \cdot  \nabla_{\!\perp}) \bm{P}$, creates flow that is perpendicular to the polarisation field and equal to its bend. This is a manifestation of the instabilities towards splay or bend in contractile or extensile active fluids, respectively \cite{Ramaswamy:2010bf}. For a drop this mechanism means that both pure splay or pure bend, in an asymmetric configuration, can generate a directed flow in the bulk of the drop and thus enable it to propel itself along the substrate. The direction of propulsion depends on the sign of the activity $\sigma_0$ and on the friction parameter $\xi$ (see Sec.\ \ref{Subsec:FrictionDependence}). Figures \ref{fig:AsterDefect} and \ref{fig:VortexDefect} show examples for orientation fields with pure splay or pure bend and the resulting flow fields. 

A polarisation field with pure splay can be produced by varying the position of an aster defect, as illustrated in Fig.\ \ref{fig:AsterDefect}. The resulting flow will be aligned with the polarisation and  have a large component in direction of the splay. However, the flow changes its direction along a line where the splay of $\bm{P}$ is balanced by the coupling term \eqref{eq:PHcoupling}, $U_h+\nabla_{\!\perp}\! \cdot \bm{P}=0$. At the right boundary the flow is driven by the vertical splay due to the tangential anchoring to the bounding surface and directed inwards in this example. Note, that if $\bm{P}$ is uniform then the coupling to the surface, that is term \eqref{eq:PHcoupling}, is solely responsible for generating a flow in the drop, which then leads to symmetric spreading studied extensively in \cite{Joanny:2012dx}. Here however, the flow at the boundary counteracts the bulk flow that is driven by horizontal variations in $\bm{P}$.
 
A polarisation field with pure bend, as in Fig.\ \ref{fig:VortexDefect}, can be generated by a variably positioned vortex defect. The flow is perpedicular to the polarisation in the bulk, having a large  component in direction perpendicular to the bend, but alignes with the polarisation close to the boundary due to the coupling term \eqref{eq:PHcoupling}. However, in the axisymmetric case in Fig.\ \ref{fig:VortexDefect}(h) $U_h$ vanishes, because $h$ is now constant along closed circles in $\bm{P}$, leading to purely radial flow. In droplets of actin and myosin solution a radially inward flow, that was associated with an emergent ring-like structure of actin, was observed experimentally \cite{2012PNAS..10911705P}. We can recover this result by choosing contractile activity ( $\sigma_0=-1$) in Fig.\ \ref{fig:VortexDefect}(h).

\begin{figure}[h]
\includegraphics[width=0.48\textwidth]{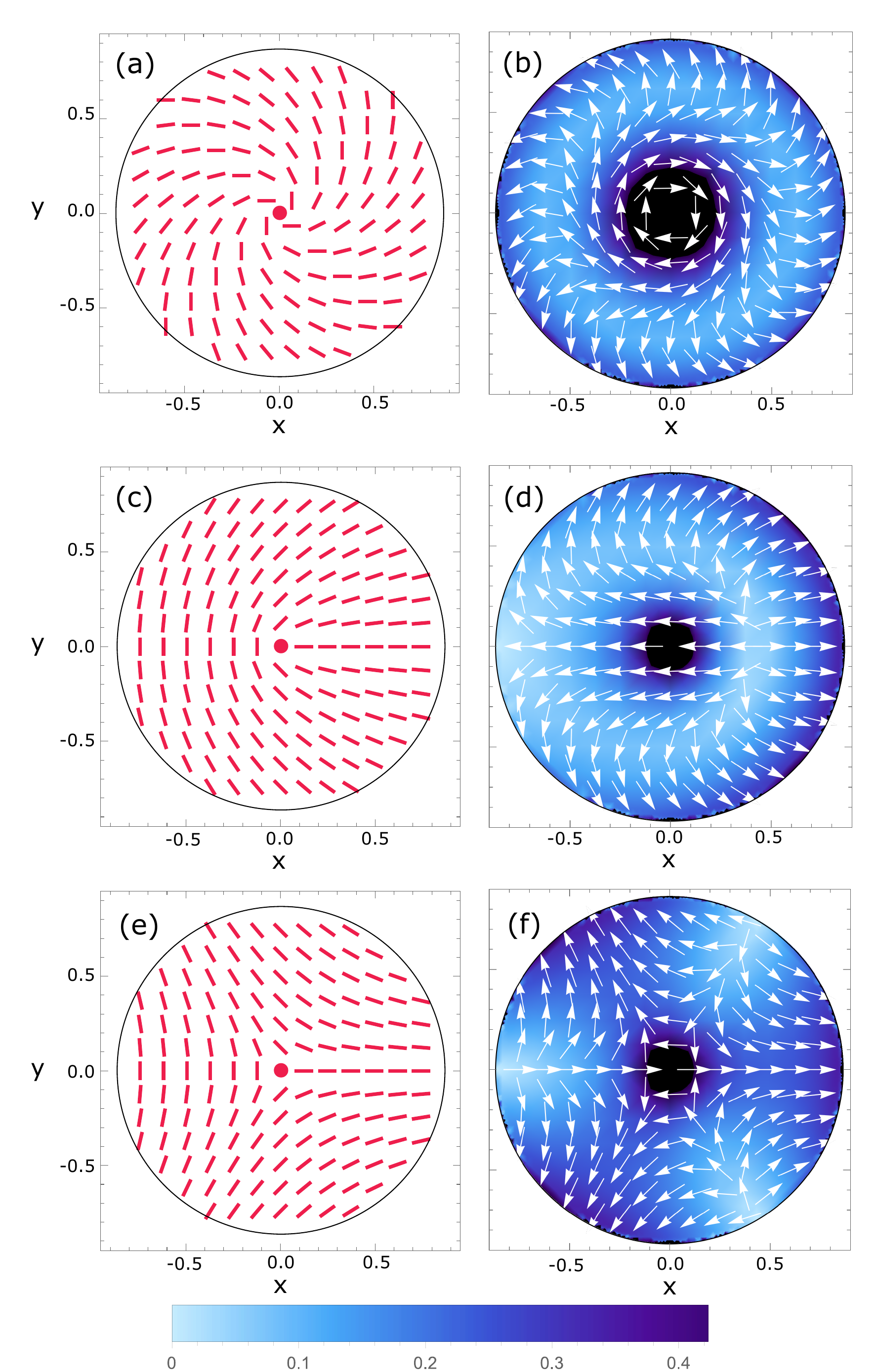}
\caption{\label{fig:DifferentDefects} {\it Left:} Top view of polarisation fields with (a) spiral ($a=\nicefrac{\pi}{4}, b=1$), (c) $+1/2$ defect ($a=0,b=\frac{1}{2}$), and (e) $-1/2$ defect ($a=0,b=-\frac{1}{2}$) in the centre of the drop ($r_0=0$). {\it Right:} Plot of the corresponding flow fields \eqref{eq:FlowDecomposition} for $z=0.01$. Parameters are $\sigma_0=1$ (extensile) and $\xi=0$ (no slip).}
\end{figure}

For more complex polarisation fields all contributions to \eqref{eq:FlowDecomposition} are present, as shown in Fig.\ \ref{fig:DifferentDefects} for three different types of defects placed in the centre of the drop. The spiral defect (Fig.\ \ref{fig:DifferentDefects}(a)) generates rotational flow around the centre. However, close to the boundary the direction of rotation reverses, which is here again a result of the coupling to $\nabla_{\perp} h$. A similar flow pattern, including the change of rotational direction, was observed in thin drops of bacteria suspensions where the swimming bacteria self-organised into a spiral vortex \cite{Wioland:2013jm}. 

In addition to profiles with integer strength defects, which can be found in systems with polar or nematic order, we also consider half-strength defects that only exist in nematic systems. Such defects emerge spontaneously and shape the dynamic flows in many active systems with nematic symmetry \cite{Sanchez:2013gt,Keber:2014fh}, particularly the self-propelling $+1/2$ defect. In our model, a $+1/2$ defect in the centre of the drop (Fig.\ \ref{fig:DifferentDefects}(c)) creates unidirectional flow in the bulk, unlike $+1$ defects in the same position, because it has an advantageous combination of a splayed and a bent region that drive the flow in the same direction. A $-1/2$ defect (Fig.\ \ref{fig:DifferentDefects}(e)) on the other hand leads to no net direction of the flow in the bulk due to the three-fold symmetry of the corresponding polarisation field. These flows are in agreement with the two-dimensional results for half defects, that explain the self-propulsion of positive and the stagnation of negative half defects in active nematics \cite{Giomi:2014ha,Giomi:2013ky}. 

A point defect in the field $\bm{P}$, when positioned in the interior of the drop, corresponds to a line defect in the field $\bm{S}$. At this line defect the magnitude of the horizontal flow typically diverges, for instance as $\sim \frac{1}{r}$ at the two $+1$ strength defects in Fig.\ \ref{fig:AsterDefect}(d) and \ref{fig:VortexDefect}(d). This could result from the fact that in the very vicinity of a defect the approximation of separable length scales breaks down, because the flow varies strongly on a small length scale in the horizontal direction, as well as in the vertical, and all spatial derivatives should remain in eq.\ \eqref{eq:NavierStokesEquation}. However, our model offers a realistic prediction for the direction of the three-dimensional flow at defect lines. In particular, the radial inflow at +1 defects is converted into the third dimension by continuity and could lead to a thin, protrusion-like deformation of the drop at the location of the defect. Such protrusions were recently observed in experiments of thin shells of active liquid crystal \cite{Keber:2014fh}, under the condition of lowered surface tension with respect to the filament elasticity, and could be the experimental equivalent of such singularities in flow that are induced by a defect.

So far we have excluded the effect of surface tension on the flow in the drop, because the associated term was of lower order in the approximation. If one chooses to include this effect and scale $\gamma \sim \varepsilon^{-3}$ in eq.\ \eqref{eq:YoungLaplace}, it will contribute 
\begin{equation}\label{eq:SurfaceTension}
\bm{u}_{\perp}^{surf} = - \left(\frac{z^2}{2}+ h \left(\xi -z\right)\right)  \gamma \nabla_{\perp} \nabla_{\perp}^2 h \,
\end{equation} 
to the horizontal component of the flow. For a no-slip boundary, this would yield an additional radially inward flow and thus either reduce or enhance the flow at the boundary due to \eqref{eq:PHcoupling}, depending on the sign of $\sigma_0$. Thus, the bulk flow would still be primarily determined by variations in the horizontal orientation profile.

\subsection{Surface friction controls speed and direction of motile drops}\label{Subsec:FrictionDependence}

\begin{figure}[t]
\includegraphics[width=0.48\textwidth]{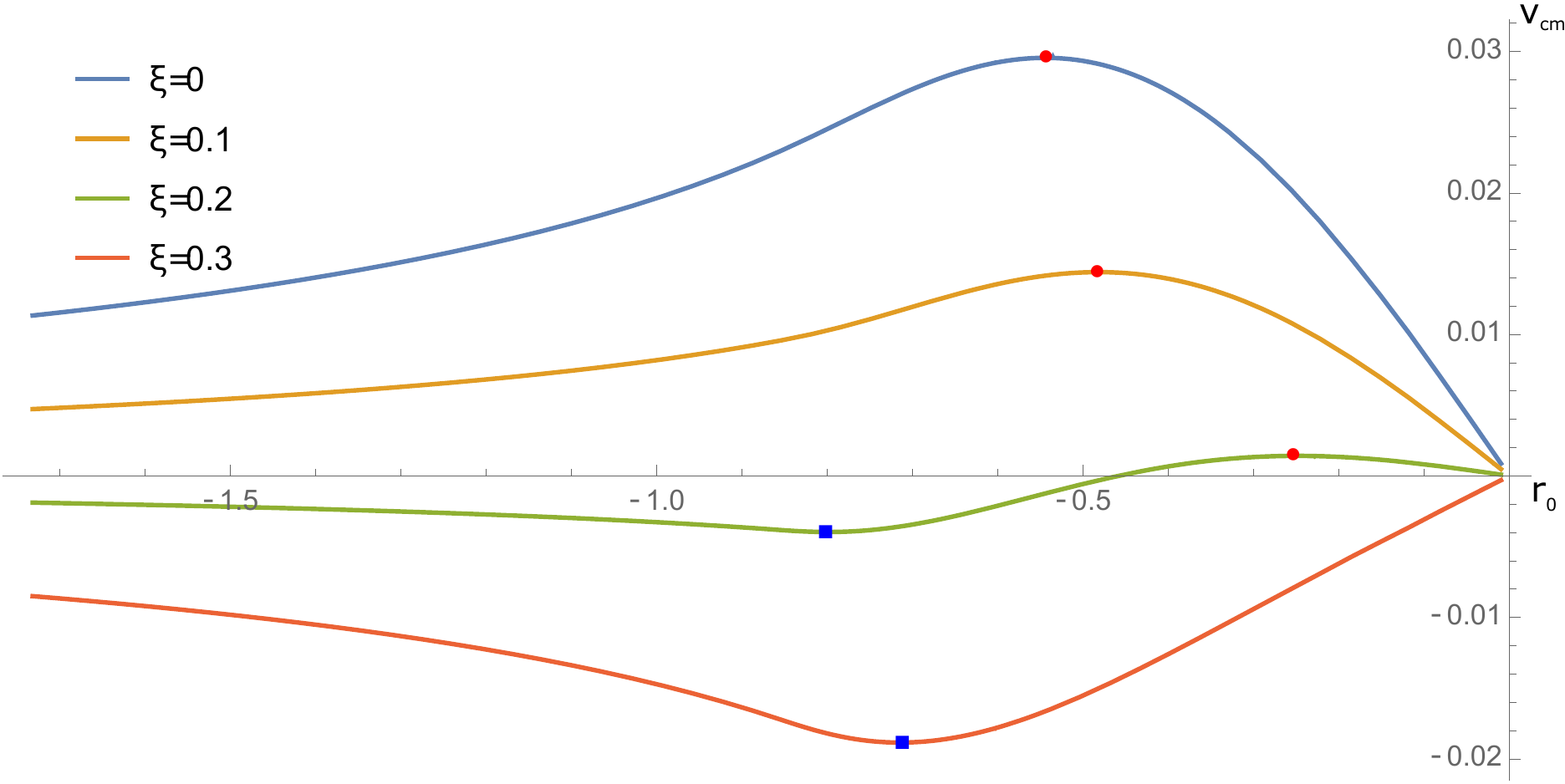}
\caption{\label{fig:Velocity} Centre-of-mass velocity for different values of friction and different positions of an aster defect in a contractile drop (or of a vortex defect in an extensile drop, respectively), with marked maxima (red dots) and minima (blue squares). A no-slip boundary yields the highest maximum velocity in the direction away from the defect and for large enough slip, $\xi \gtrsim 0.2$, the direction of propulsion is reversed. Maximal propulsion speeds are achieved with the defect being placed asymmetrically in the interior of the drop, $-1<r_0<0$.}
\end{figure}

To determine whether a drop, which produces a directed horizontal flow in the bulk (like in Fig.\ \ref{fig:AsterDefect}(f) or \ref{fig:VortexDefect}(f)), will move its centre of mass it is necessary to investigate the vertical flow component. The amount of slip at the rigid surface, represented here by the effective friction parameter $\xi$, controls whether the flow in the bulk of the drop is laminar or rotational, and in the former case determines the direction of the flow. 

Thus, surface friction has implications for the direction and speed of the centre-of-mass movement of the drop, which is summarised in Fig.\ \ref{fig:Velocity} for the example of a splayed orientation field in a contractile drop. As a measure for the self-propulsion of the drop along the substrate we numerically calculate the centre of mass velocity $v_{\textrm{cm}}$ as the integral of the flow component in $x$-direction, $v_{\textrm{cm}} = \frac{1}{V_0}\int_{\textrm{drop}} \! u \,\mathrm{d}V $. This velocity depends on the friction and the defect position, and reaches a maximum when the defect is located inside the drop. Exactly the same plot holds for a bent orientation field in an extensile drop. Surface friction is essential for the self-propulsion, since a no-slip boundary yields the highest possible maximum velocity, while an increasing slip brings the drop to a halt and, for $\xi \gtrsim 0.2$, reverses the direction of propulsion. 

\begin{figure}[t]
\includegraphics[width=0.48\textwidth]{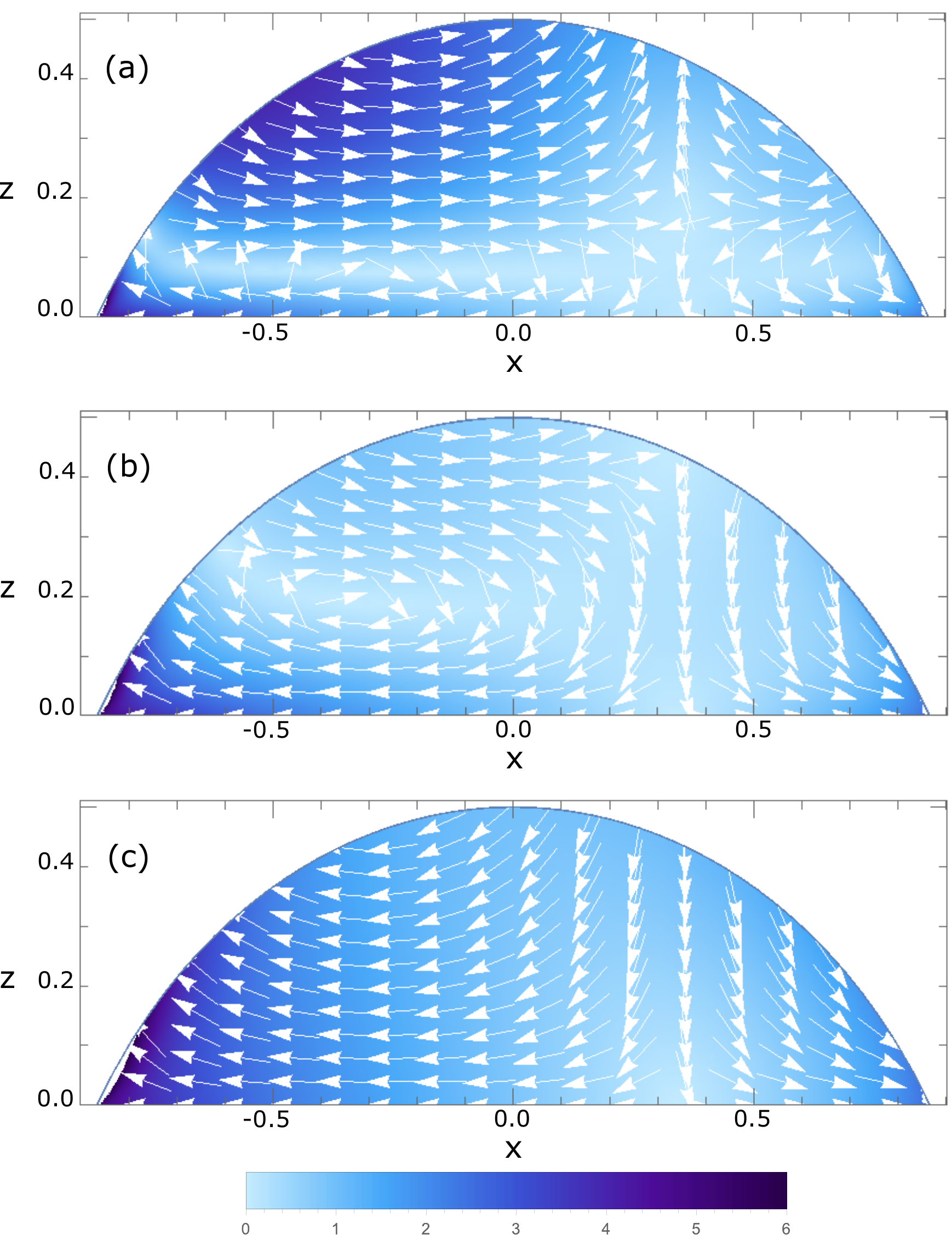}
\caption{Plot of the flow components $(u,w)$ for $y=0$, given by \eqref{eq:FlowDecomposition} and \eqref{eq:FlowVertical}, for the splayed polarisation field in Fig.\  \ref{fig:AsterDefect}(b) in a contractile drop ($\sigma_0=-1$). In this side view the defect is located in the left corner. The slip increases from (a) to (c), $\xi=0.07,0.15,1.5$. The color-coded magnitude of the flow corresponds to (c) and scales in the same way as $\xi$. \label{fig:SplayFriction}}
\end{figure}

We illustrate the effect of friction on the flow by the example of the splayed orientation field from Fig.\ \ref{fig:AsterDefect}(b), where an aster defect is located on the contact line. For a no-slip boundary ($\xi=0$) the flow vanishes at the base and, at finite height, is directed away from the defect in a mostly laminar way in the bulk of the drop (not shown). For small slip (Fig.\ \ref{fig:SplayFriction}(a)) a thin treadmilling layer emerges close to the base, on top of which the flow is still laminar. Friction with the substrate induces a shear flow that opposes the flow generated by active stresses. For finite slip this creates a vortex in the fluid that moves upwards and spans a larger region of the drop with increasing slip (Fig.\ \ref{fig:SplayFriction}(b)). If the friction is small enough, the flow becomes laminar again but with reversed direction compared to the no-slip case (Fig.\ \ref{fig:SplayFriction}(c)). 

This behaviour is apparent from the role of the friction parameter in the $z$-dependent prefactor in \eqref{eq:FlowDecomposition}, which is bounded from below and above by
\begin{equation}
h \left(\xi -\frac{h}{2}\right) \leq \frac{z^2}{2}+ h \left(\xi -z\right) \leq h \xi\,\,.
\end{equation}
While the upper bound is $h \xi \geq 0$, the lower bound becomes negative in those regions of the drop where 
\begin{equation}\label{eq:FrictionCondition}
\xi < \frac{h}{2} \,\,.
\end{equation}  
Both horizontal flow components $u$ and $v$ then change sign at a height $z_0(x,y)= h-\sqrt{h(h-2 \xi)}$. For high enough friction condition \eqref{eq:FrictionCondition} is satisfied in the bulk of the drop and the flow is rotational. In the low friction regime the transition from rolling to spreading upon increasing slip is analogous to what is found for passive fluid drops on an inclined surface \cite{Thampi:2013hm}, where a longer slip length corresponds to more sliding and less rolling of the drop. 

Fig. \ref{fig:SplayFriction}(a) represents a realistic scenario for the bulk of a polarised and motile cell extract of actomyosin enclosed by a membrane \cite{Verkhovsky:1999fe}, where the effective friction is mediated by focal adhesions that attach the actomyosin gel and membrane to the substrate. Here the bulk of active fluid flows in direction of the splay, away from the defect, thus such a cell extract would move to the right. The backslip of membrane at the substrate could be accounted for by a reduced number of focal adhesions. This flow field is consistent with the forward flow of cytosol observed experimentally in a moving cell viewed from above \cite{Keren:2009gn} and with with numerical results for the three-dimensional flow in a crawling cell  \cite{2015NatCo...6E5420T}. 

The backward flow at the leading edge in Fig.\ \ref{fig:SplayFriction}(a) (see Fig.\ \ref{fig:AsterDefect}(f) for top view) is a result of the strong tangential anchoring of filaments to both bounding surfaces, which splays the filaments vertically. This flow enhances spreading of fore-aft symmetric active droplets \cite{Joanny:2012dx}, but in the case of a strong directed bulk flow driven by horizontal variations in the orientation, as in Fig.\ \ref{fig:SplayFriction}, it can turn into a backflow on one side and counteract drop propulsion. There are two ways in which our model could be modified to eliminate this backflow at the leading edge. Firstly, we could locally remove the tangential anchoring condition in a region opposite the defect and with it the source of the backflow. Secondly, we expect that including self-propulsion of the filaments along their direction of orientation, which is the simplest way to model actin treadmilling \cite{2012PNAS..10912381T,2015NatCo...6E5420T}, would enhance the bulk flow and compensate the inward flow at the frontal boundary.

\subsection{Shape deformations}\label{Subsec:ShapeDeformation}
 
\begin{figure}[t]
\includegraphics[width=0.48\textwidth]{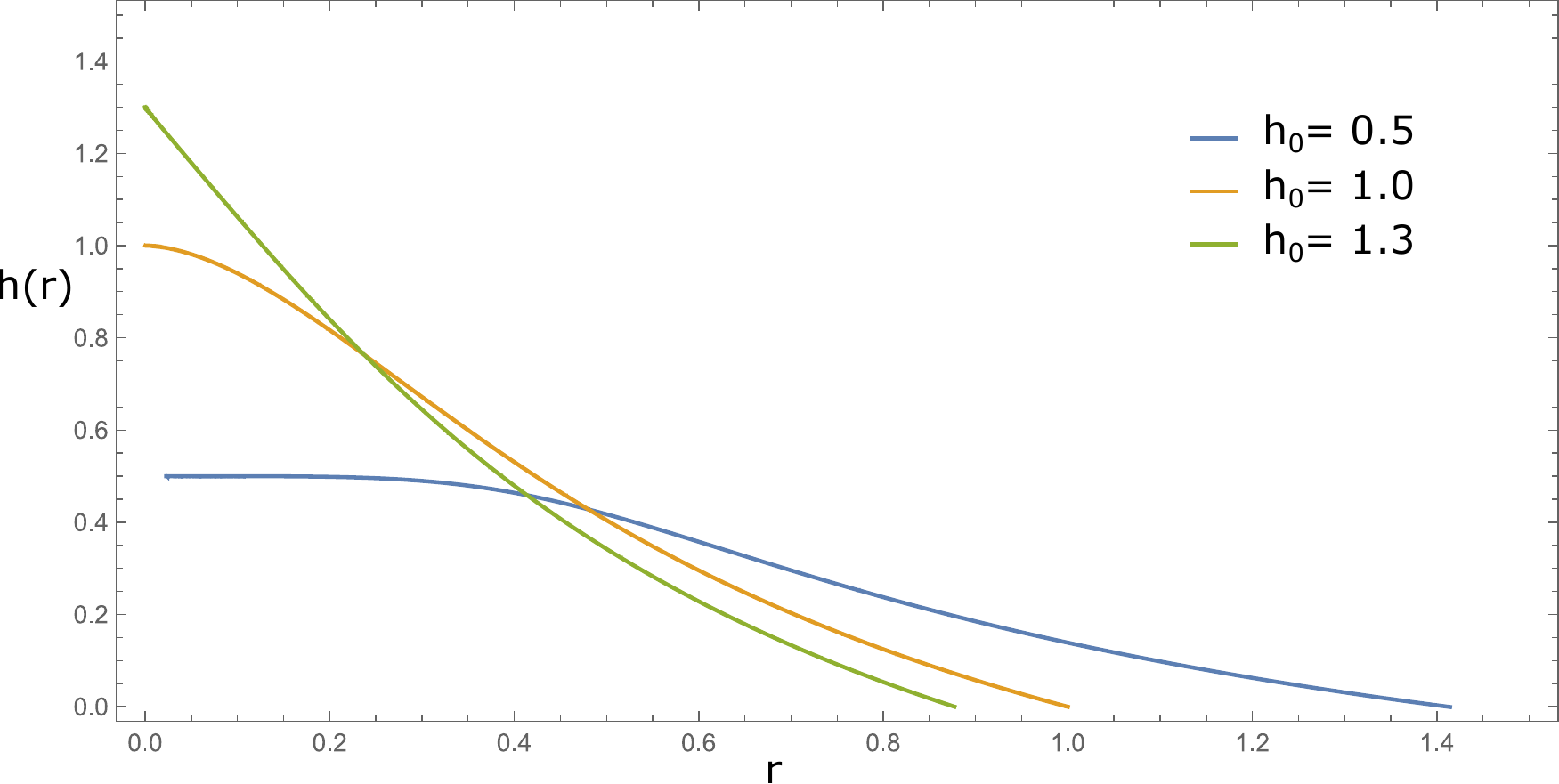}
\caption{Stationary shape profile $h(r)$ for an axisymmetric drop of unit volume with an aster defect in the centre (see Fig.\ \ref{fig:AsterDefect} (d)) and three different final heights $h_0<3\xi$, $\xi=1$. \label{fig:SteadyStateShape}}
\end{figure}

The condition of free tangential stress at the upper bounding surface means that this boundary is free to deform in response to the active flows pushing against it. The changes in shape can be expressed in terms of the horizontal flow components (see eq.\ \eqref{eq:BCkinematic2}), which themselves depend on the shape. The resulting self-consistent equation for the height function allows to not only find the immediate shape deformations due to a particular orientation field $\bm{P}$, but also to look for steady state solutions where the shape is adjusted such that the given $\bm{P}$ does not induce any deformations of the drop. 

In the axisymmetric case of an aster defect line in the centre of the drop (Fig.\ \ref{fig:AsterDefect}(d)) the time evolution of the shape is given by
\begin{equation}\label{eq:TimeEvolutionRadial}
\partial_t h  = \sigma_0 \frac{1}{r} \partial_r \left( \left(\xi -\frac{h}{3}\right)h  \partial_r \left( r h \right)  \right)  \,\,.
\end{equation} 
For finite slip, $\xi >0$, this can be solved to give an exact, implicit solution for the stationary shape of a drop of radius $r_d$ and height $h_0$ at the centre. 
The constraint of finite volume is approximated here by the condition $h_0 r_d ^2=1$. The radial dependence of the axisymmetric steady state profile is shown in Fig.\ \ref{fig:SteadyStateShape} for small heights $h_0<3 \xi$.  In particular for small values of $h_0$ our solution is very similar to the ``flat pancake'' shape that was obtained numerically in \cite{Joanny:2012dx} for small pressure. The contact angle $\theta$ for drops with $h_0<3 \xi$ is given by
\begin{equation}\label{eq:contactangle}
\tan \theta = \frac{2h_0 ^2 (3\xi -h_0)}{h_0 + 3\xi + \sqrt{3(3\xi^2+h_0(2\xi-h_0))}}\,. 
\end{equation}
It takes the maximum value of $\theta_{\textrm{max}}=\pi/4$ for $h_0=2\xi$, which also marks the cross-over from rotational to laminar flow in the large slip regime (cf.\ eq.\ \eqref{eq:FrictionCondition}).
 For $h_0 > 2 \xi$ the shape function $h(r)$ develops a singularity at the origin and for $h_0 > 3 \xi$ the solution we get is not consistent with a finite drop, indicating a lower limit for the amount of friction required to obtain a stationary shape profile. 

\begin{figure}[t]
\includegraphics[width=0.48\textwidth]{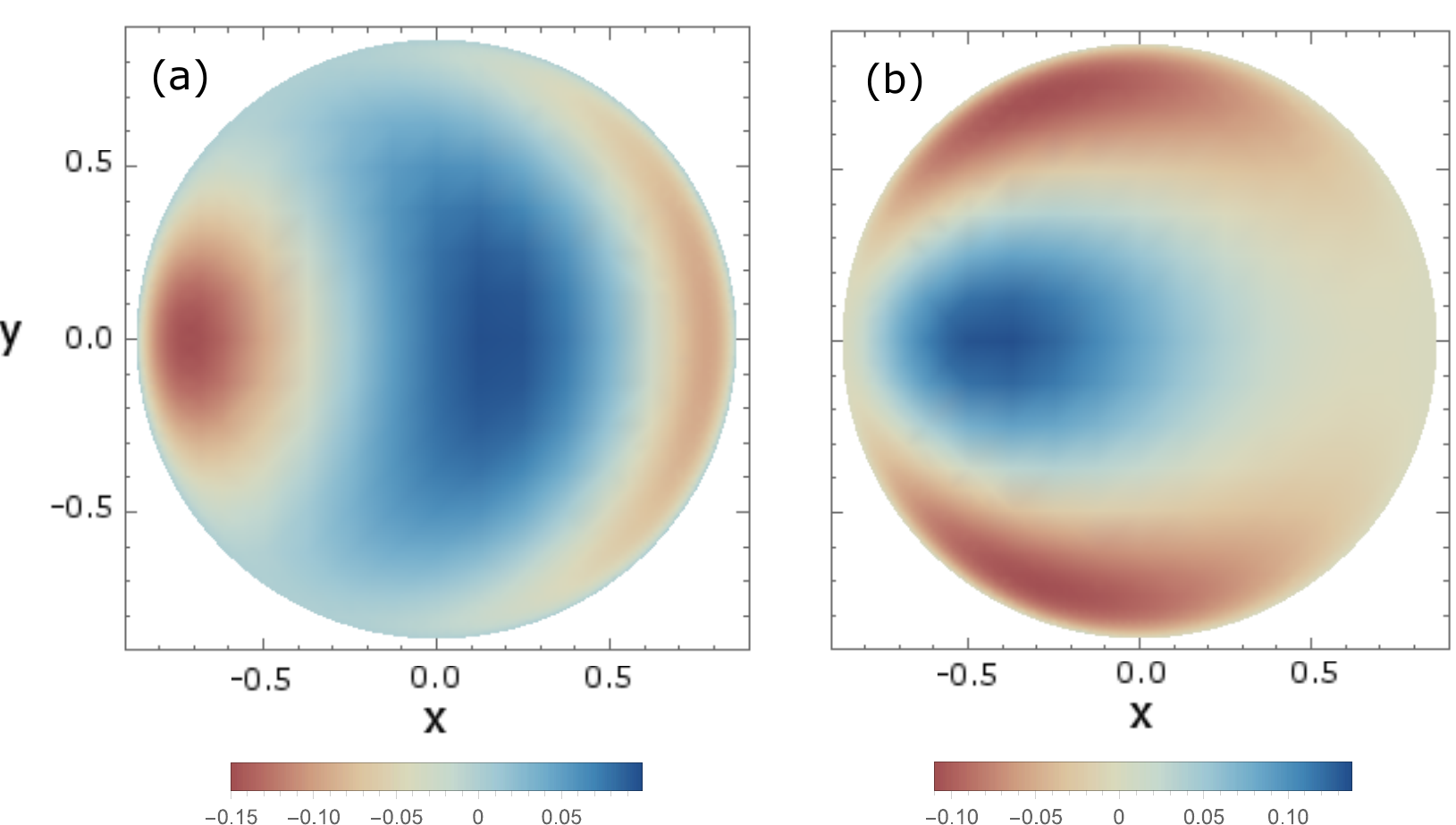}
\caption{\label{fig:ShapeDeformations} Shape deformation away from the spherical cap in the units of a small time step $\Delta t$ for a contractile drop with (a) splayed and (b) bent orientation field (see Fig.\ \ref{fig:AsterDefect}(a) and Fig.\ \ref{fig:VortexDefect}(a)) and a no-slip boundary.}
\end{figure}

We note that the steady state shape for a vortex defect line (Fig.\ \ref{fig:VortexDefect}(d)) in the centre of the drop is given in \cite{Joanny:2012dx} as a non-monotonic function with a minimum at the origin. We obtain a flow field that is consistent with this shape for an extensile suspension with a no-slip boundary, if we include the surface tension (see eq.\eqref{eq:SurfaceTension}). In this case the defect generates radially outward flow, which is opposed by an inward flow at the boundary due to surface tension, suggesting that the fluid will accumulate along a ring.

For a general polarisation field $\bm{P}$ the drop shape evolves according to
\begin{equation}\label{eq:TimeEvolution}
\partial_t h = \sigma_0 \partial_i \left( \left(\xi -\frac{h}{3}\right)h \partial_j \left(h P_i P_j \right)  \right)\,.
\end{equation}
For the asymmetric polarisation fields considered in \ref{Subsec:PolarisationDependence} it is not possible to obtain an analytical result for the steady state. We can however qualitatively describe the deviation $\Delta h$ from the initial spherical cap shape at time $t_0$ after a small time interval $\Delta t$ if we plot $\Delta h \approx \partial_t h (t_0) \Delta t$ for a particular orientation field (Fig.\ \ref{fig:ShapeDeformations}). We consider a contractile drop a with no-slip boundary for this purpose. For a splayed orientation field as in Fig.\ \ref{fig:AsterDefect}(a) the instantaneous flow results in an increase in height along a crescent-like area and a decrease to the front and back of it (Fig.\ \ref{fig:ShapeDeformations}(a)). For a bent orientation field as in Fig.\ \ref{fig:VortexDefect}(a) the height grows in the rear of the drop and descreases along the sides, so we can expect the drop to become more slender. However, the changes in shape will alter the orientation field and thus the flow. To investigate the dynamics of the shape over longer time periods it will be necessary to perform numerical simulations, which is not in the scope of this paper.

\section{Discussion}\label{Sec:Discussion}

We have given an analytic description of dynamic steady state motion of active drops. In the scope of a thin drop approximation, we derived exact expressions for the flow in the drop that is generated by a given orientation profile. This leads to the identification of two key requirements for the self-propulsion of thin droplets of active fluid on planar substrates. These are an asymmetrically splayed or bent orientation field, for instance induced by a topological defect in the interior of the drop, and sufficient surface friction provided by the substrate for the drop to push itself forward. The active flows in the drop are primarily driven by horizontal variations in the orientation and the coupling to the shape contributes mainly at the contact line, where it can generate flows that work against drop propulsion. A motile drop will become stagnant, treadmilling on the spot, when the surface friction drops below a certain value. A substrate with a large slip length could produce drops that move in a direction opposite to the expected, which is in the direction of splay for contractile cell fragments \cite{Verkhovsky:1999fe} or in direction of the comet-head for drops of extensile active nematics with a positive half-defect \cite{Sanchez:2013gt}. 

Our model suggests that the flow of cytosole in a motile cell fragment can result from active stresses in the polarised and splayed actin cytoskeleton. The result is in qualitative agreement with the forward directed flow measured experimentally in rapidly moving keratocytes \cite{Keren:2009gn}. Although the role of fluid flow in the crawling machinery is still unclear \cite{Lenz:1338869}, it was suggested to be responsible for active transport of actin monomers to the leading edge \cite{2003Sci...300..142Z}. 

Shape deformations of active droplets can be studied experimentally on vesicles filled with an active suspension, where the surface tension can be reduced by applying hypertonic stress \cite{Keber:2014fh}. To impose a particular, static orientation of active filaments the drop could be placed on a micropatterned substrate \cite{Vignaud:2012be}, specifically to induce axisymmetric defect configurations which would allow to observe the corresponding stationary shapes of the drop. 


%





\begin{acknowledgments}
This work was supported by the UK EPSRC through Grant No. A.MACX.0002. 
\end{acknowledgments}

\bibliography{flatdrop_paper}

\end{document}